# EFFECT OF LASER POLARIZATION ON ATOMIC AND IONIC EMISSIONS IN LASER INDUCED BREAKDOWN SPECTROSCOPY (LIBS)

Adarsh U. K.[a], Unnikrishnan V. K.[a, b*], Parinda Vasa[c], Sajan D. George[a, d], Santhosh Chidangil[a, b] and Deepak Mathur[a]

[a] *Department of Atomic and Molecular Physics, Manipal Academy of Higher Education, Manipal 576104, Karnataka, India*

[b] *Centre of Excellence for Biophotonics, Manipal Academy of Higher Education, Manipal 576104, India*

[c] *Department of Physics, Indian Institute of Technology, Mumbai 400076, India*

[d] *Centre for Applied Nanosciences (CANs), Manipal Academy of Higher Education, Manipal 576104, India*

[*] Corresponding Author

E-mail address: unnikrishnan.vk@manipal.edu (Unnikrishnan V.K.)
ORCID: https://orcid.org/0000-0003-3907-0586

## Abstract

Laser induced breakdown spectroscopy (LIBS) has become a proven contemporary workhorse for qualitative and quantitative analysis of materials. Recent developments in LIBS have been limited either to signal enhancement strategies or to advances in data analysis techniques that yield better interpretation of LIBS data. Explorations of the initial laser excitation stage of LIBS remain somewhat restricted. In particular, the influence, if any, of the polarization state of excitation laser remains unexplored to a large extent. The current work addresses this lacuna in knowledge by probing the influence of different polarization states of the excitation laser on LIBS spectra of metallic copper. Specifically, we investigate the behaviour of atomic and ionic emission lines with respect to change in polarization state of the incident laser light; our results show distinct polarization dependence. Our observations open up new opportunities of tackling the problem of relatively faint emissions from ionic species in the plasma by adjusting the polarization state of the laser. Our findings also highlight the urgent need for appropriate theoretical study to be undertaken so that proper insights can be developed into the physics that drive the observations that we report here.

# 1. Introduction

Laser-Induced Breakdown Spectroscopy (LIBS) has developed into an established optical technique that is widely applicable in qualitative and quantitative analyses of materials [1,2] based on elemental composition. The technique is often the preferred tool for trace element detection [3,4]. The wide acceptance of LIBS in the field of material analysis is based on its characteristic properties like in-situ analysis, stand-off capability, low limits of detection, universal applicability, and minimal (micron-scale) destruction of the sample. The LIBS technique principally creates a micro-plasma on the sample surface by ablating the material under study; this is realized by focusing a high-energy, pulsed laser beam at the target [5,6]. Excitation of the thus-formed plasma, followed by subsequent de-excitation, gives rise to emission of characteristic spectral lines of the elemental constituents of the irradiated sample. These lines may result from either the neutral (atomic) species in the plasma or from sample constituents which have become ionized. Emissions from small molecular species in the material may also contribute towards the overall emission. Collection and analysis of the signal emitted from the irradiated material enables identification of the elements present in the sample [7], and facilitates detection of the presence of foreign elements in trace amounts. LIBS has begun to find utility in industrial applications, for quality control [8], and for material identification and classification [9,10]. The majority of contemporary reports of LIBS appear to focus on new strategies for signal collection so as to meet requirements of a particular and specific application, and to enhance the sensitivity of measurements [11-13]. Recently, much attention has been paid to the signal analysis part, with deployment of new and varied statistical tools and multi-variate data analysis techniques [14, 15].

It is now an established wisdom that the excitation part of the LIBS process is driven by experimental parameters like excitation laser wavelength, laser energy, focal length of the lens used for focussing, laser pulse width and the laser's repetition rate. These parameters have consequences on laser spot size, peak power, and fluence at the sample surface. Additionally, time-dependent studies require use of a gated detector; parameters like gate width, gate delay, and other acquisition parameters have to be thoroughly optimized prior to the application. There exist a number of cogent reports on the effect of all of these parameters on the overall LIBS phenomenon and approaches for optimizing these parameters [16,17, and references therein]. Comparison of laser-induced ablation of the sample using nanosecond and femtosecond laser pulses, with corresponding differences in peak power, has also been well described [18, 19]. However, an important parameter that seems to have attracted very little attention of the LIBS community is the polarization state of the excitation laser beam, and this is the focus of our attention in the present study.

In general, a LIBS practitioner has the option of utilizing a laser beam that is in one of three easily accessible states of polarization: linearly polarized, elliptically polarized, or circularly polarized. The output from the overwhelming majority of standard laser sources is a linearly polarized beam. It is readily possible to achieve the other two polarization states by using waveplates such as a halfwave plate or a quarter waveplate [20]. Earlier reports on polarization dependent effects that are pertinent to LIBS are limited to (i) examining the effect of polarization state on the basic photoionization process [21,22], (ii) probing changes in plasma emission that occur upon switching from linearly to circularly polarized excitation beams [23], and (iii) polarization characteristics of the fluorescence signal that is collected [24]. However, systematic studies of the role of all possible states of polarization of the laser beam in a LIBS experiment are yet to be investigated.

We report here systematic experimental investigations into the behaviour of atomic and ionic emission lines obtained upon irradiation of Cu with ~250 mJ worth of irradiation by 10 ns long pulses of 532 nm laser light with respect to change in polarization state of the incident laser light. The results that we present here show distinct polarization dependence and, in the absence of proper physical insights, serve to highlight an urgent need to develop appropriate theoretical work to be undertaken so that proper insights can be developed into the physics that drive the observations that we report here. On a practical level, our morphological findings open up new opportunities of tackling the problem of relatively faint emissions from ionic species in the plasma by adjusting the polarization state of the laser.

## 2. Methodology

### 2.1. Experimental

The experimental system used for our study comprises a conventional LIBS setup in back-collection geometry, as is schematically depicted in Figure 1. The second harmonic of a nanosecond Nd:YAG pulsed laser (Q-Smart 450, Quantel, France) is used as the source of laser light. The temporal width of our pulses is 6 ns, with a repetition rate of 10 Hz. The maximum energy at the output wavelength of 532 nm is 225 mJ. For data acquisition, we utilized a high-resolution Echelle spectrograph (Mechelle, ME5000, Andor, Ireland) coupled with an Intensifier Charge Coupled Device (ICCD) (Andor iStar, DH734-18U-03PS150 Andor, Ireland). The output from the laser source is directed towards a high energy threshold, 532 nm non-polarizing type beam splitter (50:50 splitting) kept at $45^0$ with respect to the incident beam (Fig. 1); this directs the laser beam to a UV-grade quartz lens (1-inch aperture, 10 cm focal length) which focuses the laser light on to our sample which is mounted on a translation stage. Prior to the beam splitter, we introduce a half-wave plate (or a quarter-wave plate) in order to alter the laser light's polarization state which is initially measured to be linearly polarized.

Signals emitted by the sample are collected using the same lens and are collimated onto a second lens of the same material and dimensions but with a focal length of 5 cm. This second lens focuses the signal on to the input of an optical fibre cable (200 μm core diameter). The optical fibre carries the signal into the spectrograph-ICCD system. The ICCD is gated with respect to the laser pulses using a digital storage oscilloscope interfaced to a laboratory computer.

The sample used for the present series of experiments is a high-purity (99.9 %) copper strip which is mounted on a clamp fixed on the translation stage. During the data acquisition, the translation stage is made to move in the X and Y directions so as to ensure that a fresh sample surface area is presented to subsequent laser pulses for data acquisition. On the basis of earlier studies on Cu carried out in our laboratory [17] the optimised acquisition parameters for gate delay and gate width that we used was 900 ns and 6 μs, respectively.

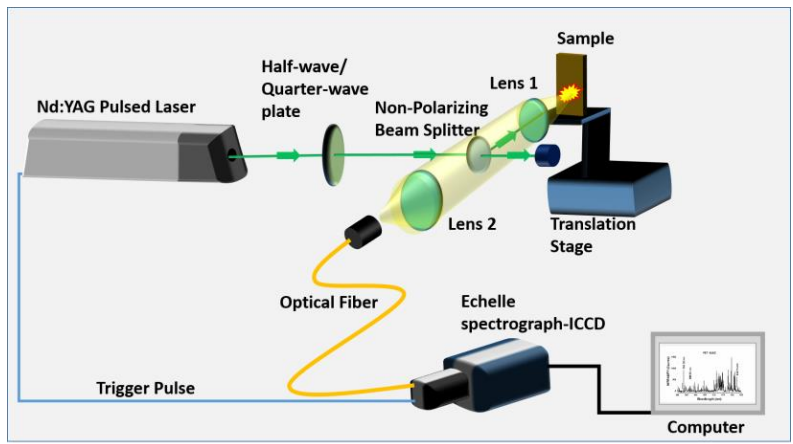

**Figure 1**: Experimental system used in the present set of LIBS experiments.

The initial experimental task is to accurately determine the direction of linear polarization at the laser output. We utilized a high energy threshold, reflecting type thin film polarizer, which reflects S-polarized light and transmits P-polarized light when it is kept at its specific Brewster angle ($56^0$). The thus-determined polarization direction was used as the reference direction with which the axes of the wave plates were aligned at the beginning of each series of measurements.

LIBS spectra of copper were measured after introducing wave plates that induced changes in the polarization state incident on the sample. The half-wave plate does not alter the linear polarization but induces a change in direction of the polarization axis of 2θ if the axis of the half-wave is oriented at an angle θ with respect to the direction of polarization of the incident beam. On the other hand, the quarter-wave plate oriented at $45^0$ with respect to the polarization direction of the incident beam induces a change in the polarization state, making the initial linearly polarized beam into a circularly polarized one. If quarter-wave plate is at any angle between $0^0$ and $45^0$, the corresponding transmitted beam will be elliptically polarized.

We investigated changes in the LIBS spectra of Cu when the polarization direction of the linearly polarized excitation laser beam is altered from the inherent direction. For each $20^0$ rotation of polarization angle (with respect to the original direction), covering the range $0^0$-$180^0$, five spectra were acquired. Similarly, upon introducing the quarter-wave plate in the laser path, five spectra were acquired by keeping the axes at $0^0$, $20^0$, $45^0$, $70^0$, and $90^0$ with respect to the polarization direction of the laser.

A critically important facet of our experimental measurements was to thoroughly rule out the possibility of external factors other than polarization state and the polarization direction of the excitation laser beam being responsible for changes (if any) in the LIBS spectra. To monitor changes due to possible pulse-to-pulse variations in the laser energy during the acquisition, several sets of measurements over an extended period of time were carried out. Another potential factor is possible changes in energy and beam characteristics of the excitation beam caused by the presence of wave plate in the beam path. We took care to monitor the beam characteristics of the excitation beam before and after the wave plates for different orientations of the waveplates with respect to the polarization direction of the beam. Such monitoring was performed using the moving-slit method where a slit of fixed slit width is mounted with a high energy threshold laser power meter in the same base mount. At different orientations of the wave-plates that result in different polarization directions in the case of the half-wave plate and different polarization states in the case of a quarter-wave plate, the laser beam profile was plotted, and beam width was determined. In addition, the maximum energy obtained at each stage of the laser beam was also determined from the plots as all the measurements were performed at the same input laser energy of 10 mJ. All measurements are performed at a fixed position of about 70 cm away from the laser head.

## 3. Results and Discussion

### 3.1. Observations

A typical LIBS spectra that we measured of our pure Cu sample is shown in Figure 2 along with the wavelengths of the major emission lines recorded in our experiments. These measurements were made without any wave plate in the optical path of the laser beam and, hence, pertain to a linearly polarized incident laser beam.

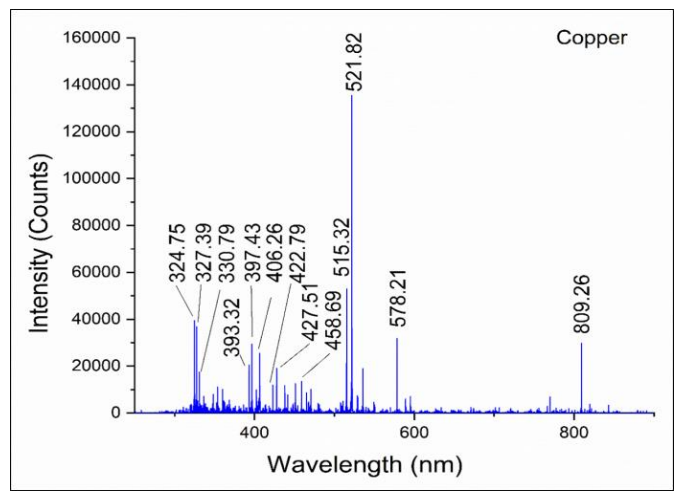

**Figure 2**: LIBS spectrum of copper acquired with linearly polarized laser light.

The prominent atomic emission lines from copper include 324.75 nm, 327.39 nm, 330.79 nm, 406.26 nm, 427.51 nm, 458.69 nm, 521.82 nm, and 809.06 nm. Ionic emission lines include 393.32 nm, 397.43 nm, and 422.79 nm. We also made measurements on the variation in LIBS spectra of

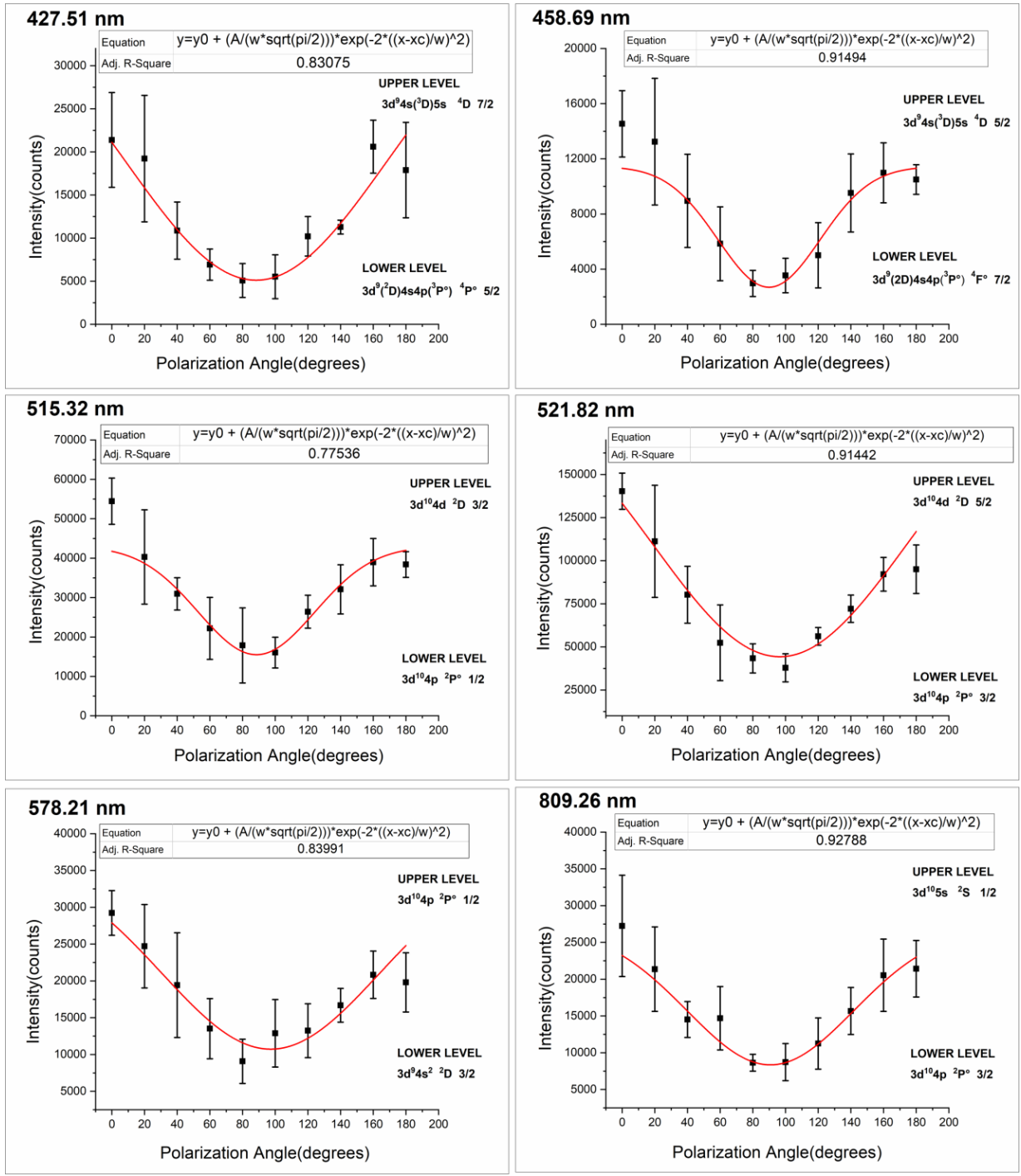

**Figure 3**: Variation in intensities of atomic emission lines in LIBS spectra of copper as a function of polarization direction of a linearly polarized excitation laser beam (all data fitted to Gauss function with equations indicated in each panel).

Cu with respect to variation in the polarization direction of our linearly polarized excitation laser beam and our results are represented as the emission line-wise variation in line intensities with respect to each $20^0$ orientation of polarization direction induced by the half-wave plate in the optical path.

Figure 3 represents variations in the atomic emission line intensities of copper spectra upon variation of polarization direction with respect to the inherent direction of polarization of the laser beam. The energy levels of Cu associated with each of the emissions are also shown. From Figure 3 it can readily be inferred that the intensity of atomic emission lines from Cu are most certainly influenced by the polarization direction of the excitation laser. The maximum intensity is observed when the polarization direction of the beam is along the inherent direction of polarization of laser (which happens to be vertical in our experimental conditions). As the polarization direction is gradually altered from vertical to horizontal, the corresponding intensity is seen to decrease. We note that the data points depicted in Fig. 3 are the average of five trials and, hence, the possibility of random fluctuations is negligible. We confirmed this by plotting the standard deviation at each angle of polarization and establishing the rigidity of the trend that is observed in all the emission lines shown in Fig. 2. The vertically polarized laser appears to induce the highest emission intensity in the LIBS spectra of Cu measured in our experiments.

We have discovered that the behaviour exhibited by the ionic emission lines from the same Cu sample with respect to the change in orientation of laser beam polarization is completely different from that observed in the case of atomic emission lines.

The results obtained in the case of ionic emission lines are depicted in Figure 4. It is seen that the maximum intensity exhibited by the ionic emission lines in the LIBS spectra of copper is not when the direction of polarization of excitation laser is vertical, as in the case of atomic emission lines. As the plots in Fig. 4 suggest, the maximum intensity is neither when the polarization direction is vertical nor when it is horizontal, but at an orientation of around $45^0$ and $135^0$ with respect to the initial direction of polarization of laser (in the data presented in Fig. 4, readings are at an interval of $20^0$ over the range $0^0$ - $180^0$). This behaviour of ionic emission to the change in the polarization direction of laser is quite unexpected and may be of utility in enhancing the ionic emissions in LIBS spectra which are usually weak compared to atomic emissions.

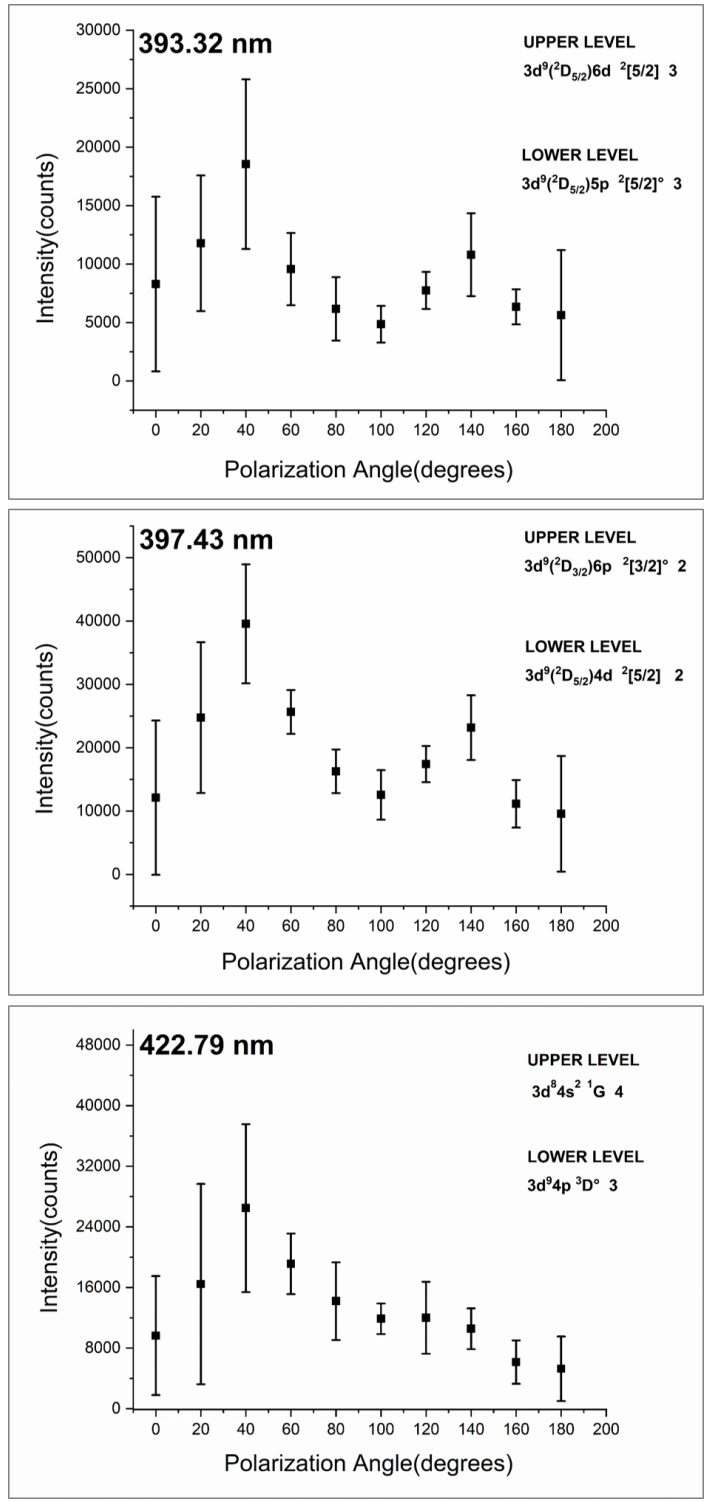

**Figure 4**: Variation in absolute intensities of ionic emission lines in LIBS spectra of copper with respect to the polarization direction of the linearly polarized excitation laser beam.

Circularly and elliptically polarized beams were realized in our experiments by replacing the half-wave plate with a quarter-wave plate in the optical path (Fig. 1). Typical observed behaviour of both atomic and ionic emission lines in the LIBS spectra of Cu with respect to the change in the polarization state of the excitation laser beam are shown in Figures 5 and 6.

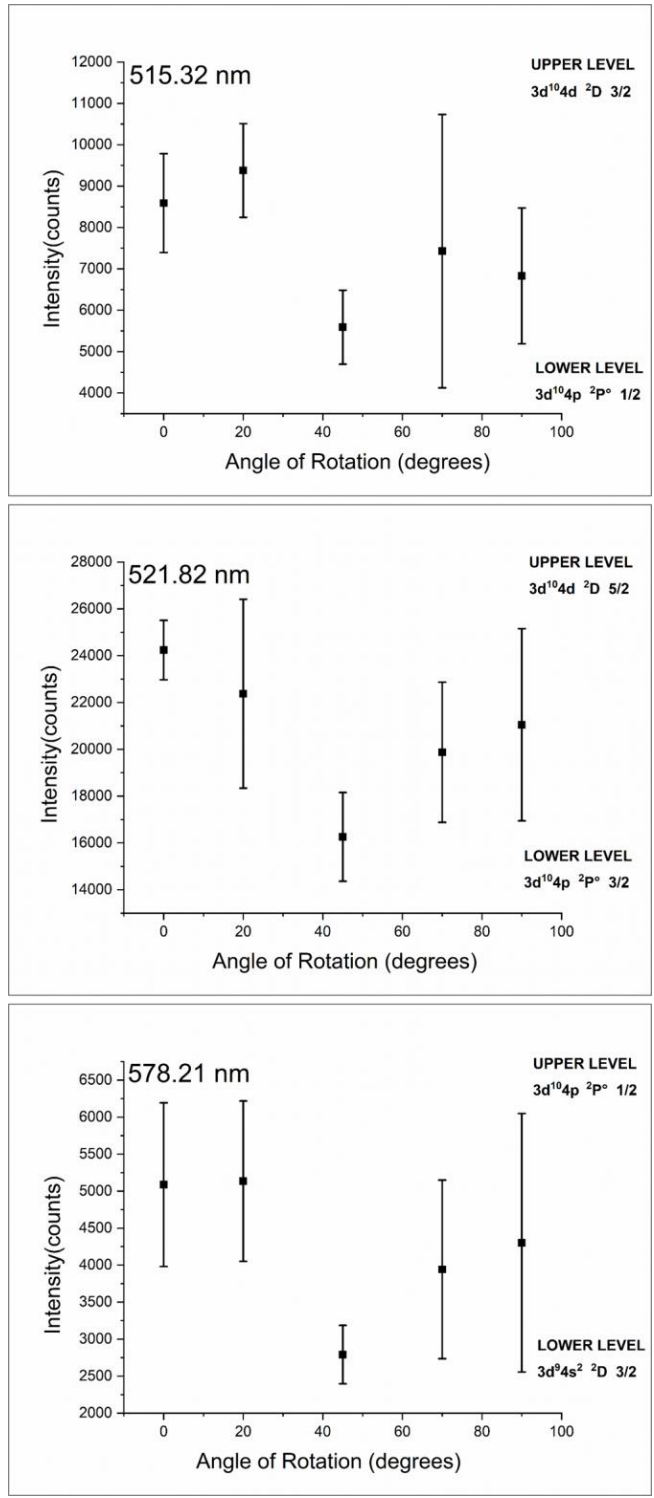

**Figure 5**: Variation in intensities of atomic emission lines in LIBS spectra of Cu with change in the polarization state of the excitation laser beam.

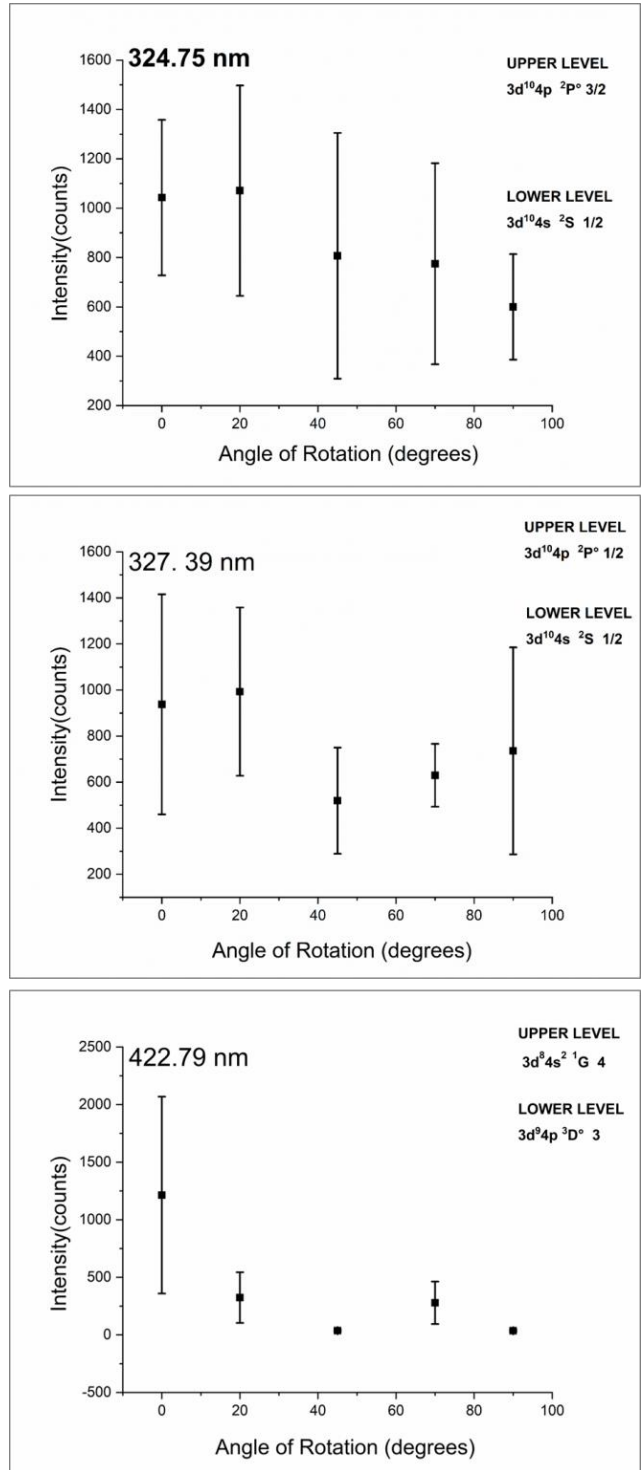

**Figure 6**: Variation in intensities of ionic emission lines in LIBS spectra of Cu upon change in the polarization state of the excitation laser beam.

It is seen that maximum emission is obtained from atomic excitation when the polarization state is linear (when the orientation of axes of the quarter-wave plate with respect to the reference direction is either $0^0$ or $90^0$). Least emission signals were recorded when the polarization was circular (when the axis of the quarter-wave plate was oriented at $45^0$ with respect to the reference direction).

Ionic emission lines also appear to behave in the same manner (Fig. 6): maximum emission is observed at $0^0$ and $90^0$ orientation of axes of the quarter-wave plate with respect to the reference direction. Minimum emission is obtained at $45^0$ orientation.

As already mentioned, we took cognizance of the possibility that the apparent polarization dependence of LIBS signals might result from other factors, like changes in laser energy and laser beam characteristics due to the presence of wave plates which, in turn, might influence the excitation-emission process. The laser beam characteristics recorded in the presence of wave plates positioned at different orientations with respect to the reference direction of polarization are shown in Figure 7(a) and 7(b), respectively. Figure 7(a) shows that both beam shape and beam width remain essentially the same when the polarization angle is at $0^0$, $90^0$, and $180^0$ with respect to the initial polarization direction.

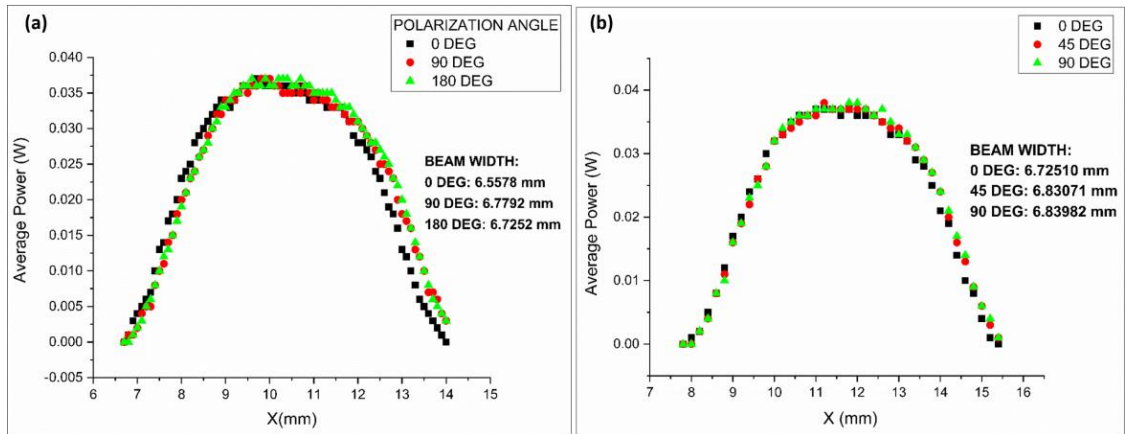

**Figure 7**: Laser beam profile in the presence of (a) half-wave plate at different orientations and (b) quarter-wave plate at different orientations.

The maximum power after the wave plates was also measured to remain constant at these positions, revealing that the presence of a half-wave plate in the optical path before excitation is not significantly influencing other factors like laser energy or laser beam shape.

Observations from Fig. 7(b) suggest that there are no evident changes in the beam profile of the excitation laser when the quarter-wave plate is introduced in beam path. Measurements were performed at three different angles of orientation of the axis of the quarter wave plate with respect to the linear polarization direction of the laser beam. The shape of the beam and the maximum power was confirmed to remain essentially invariant for all our experimental conditions, confirming that the polarization dependence of LIBS spectra is not likely to have been caused by any changes in the laser beam characteristics in the presence of wave plates.

Similarly, changes in the energy of the excitation laser beam were also monitored in the presence of half-wave plate and quarter-wave plate at different orientations; typical results are shown in Figure 8(a) and 8(b), respectively.

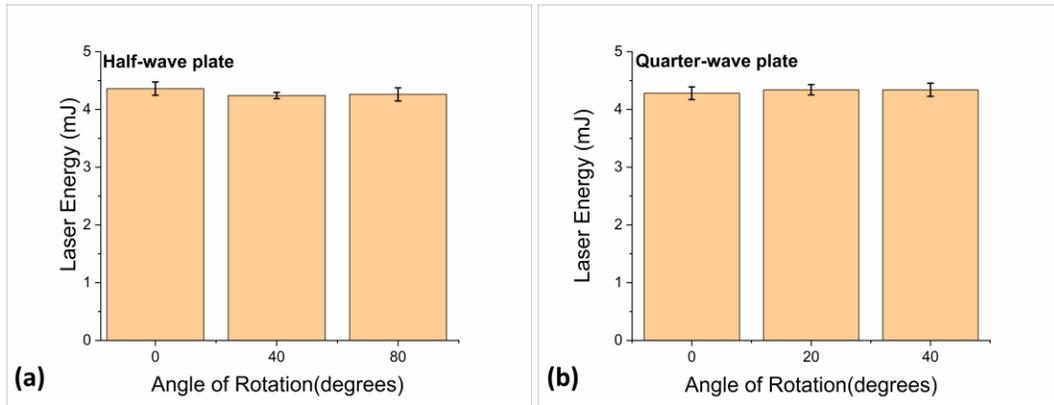

**Figure 8**: Variations in laser energy at different orientations of the (a) half-wave plate and (b) quarter-wave plate.

For a fixed laser energy (4.3±0.1mJ) measured before the wave plates, the variations in energy after the wave plates were recorded at least 5 times, with interval of 30 seconds between each measurement. Figure 8(a) represents the variations in energy when the axis of the half-wave plate was oriented at $0^0$, $40^0$, and $80^0$ with respect to the reference direction of the beam: we see no evidence of significant variations in energy that may influence the excitation process. Similarly, readings of laser energy were noted for different orientations of our quarter-wave plate, and similar results were obtained. The results also served to confirm that the emission changes that we report are unlikely to be caused by energy changes brought about by the presence of wave-plates at different orientations.

To check whether the reflective-element beam splitter (non-polarizing type/ Bk7 substrate) used in our optical system (50:50 splitting for $45^0$, Fig. 1) has any dependence on polarization of the incident laser beam, we monitored the laser energy after the beam splitter under different polarization conditions; results similar to shown in Fig. 8 were obtained.

**3.2. Discussion**

The experimental observations unambiguously shown that polarization of the excitation laser beam influences the atomic emission line intensity in laser-induced breakdown spectroscopy of metals. In the absence of proper theoretical underpinnings, we offer in the following some speculations in respect of the observations that we have reported in the hope that these stimulate appropriate theoretical work.

At a superficial level it might be presumed that our polarization-dependent results are possibly ascribable to formation of surface plasmon polariton (SPPs) at a particular polarization direction of the excitation laser beam. Since our sample is metallic and our experiments were conducted in an air environment, the general metal-dielectric condition is, in general terms, fulfilled for formation of surface plasmon polaritons. Although the propagating SPPs which are generated at the metal-dielectric interface would be expected to require a thin metal film, and would be generated for only specific angle of incidence, it is possible that generation of localized SPPs on our metal surface may be induced by surface roughness or by the expanding plasma that is formed upon irradiation. However, we have to take cognizance of the fact that surface plasmon polaritons require the dielectric constant of the irradiated metal to be negative. This would imply very high reflectivity, as would be expected in metals like silver. Copper, in the range of ~500 nm, is not highly reflective. Moreover, the dielectric constant of pure Cu in vacuum is also only very slightly negative. Both these factors would be sufficient to the supporting of plasmon polaritons a most unlikely possibility. In this context, we also note that rationalization of polarization dependency that is based solely on surface roughness would need to account for the fact that roughness would generally be expected to possesses random orientations. Further work is clearly warranted, perhaps by conducting measurements in which surface roughness is quantitatively determined and/or controlled.

Another possible scenario that merits further study is that plasma-polariton interactions might occur that lead to reduction of the emission intensity from the plasma. As the polarization direction of the laser is changed from vertical to horizontal, the magnitude of the optical field alters, which might induce an altered plasmonic interaction with the metal surface at that particular wavelength. Under these conditions, the emitted light from the plasma might couple with the plasmons, resulting in reduction of emission intensity. The influence of polarization on the electric field generated by surface plasmon polaritons in general have been reported [25] and cogently reviewed [26]; SPP localization and field enhancement has been found to be highest for a radially polarized incident optical field. When the excitation field excites the metal surface with s-polarized light, the electric field intensity is the weakest; maximum is obtained when it is p-polarized.

The interpretation of observations in the case of circularly or elliptically polarized laser excitation beam is interesting and most certainly merits theoretical efforts. It is well established from studies conducted with high-intensity laser beams [27] that the shape of the usual Coulomb potential function describing an atom changes when linearly polarized light is replaced by circularly polarized light. In the latter case it assumes a doughnut-like toroidal shape with a saddle point at the atomic position instead of the Coulombic singularity [27]. For high enough values of laser intensity (high values of optical field strength), the electric field of the circularly polarized light induces the atomic electrons in a circular trajectory in the plane of polarization, with the circle being centered at the atomic nucleus. This is schematically depicted in Figure 9. Explorations of how properties such a pulse profile and polarization state of the incident high-intensity laser fields affect the ionization dynamics of atoms and molecules have been widely reported in the course of the last few years (see [28,29] and references therein).

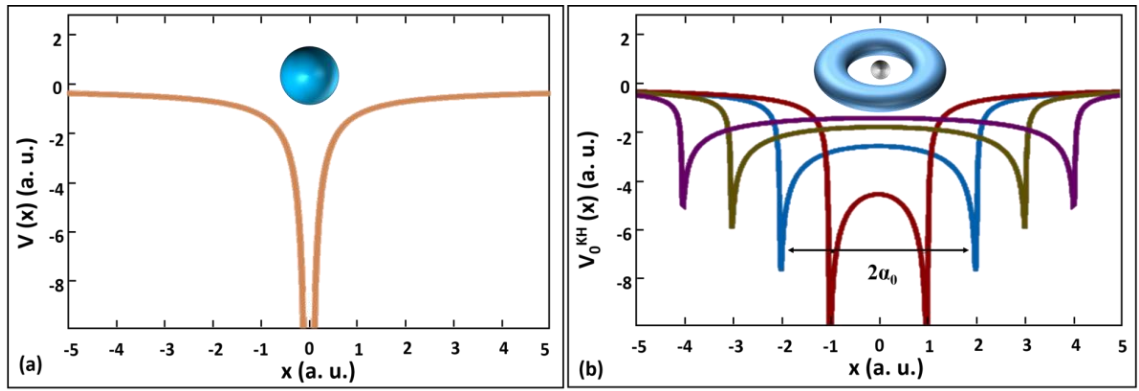

**Figure 9:** Schematic depiction of changes in a single-electron atom upon irradiation by (a) linearly polarized light and (b) circularly polarized light represented in terms of radial potential functions.

It is expected that, under conditions depicted in Fig. 9 (b), ionization would be reduced as would excitation. In the case of extremely weak laser fields used in our experiments, it is only the excitation process that would occur, and the different morphologies of the electronic charge distribution in the vicinity of the irradiated atom would be expected to result in differing excitation dynamics. The nonlocalized nature of the electron density distribution under conditions of circular polarization should be consistent with the notion that the propensity for excitation (and ionization in stronger laser fields than used in the present study) is reduced compared to when irradiation is with linearly polarized beams.

## 4. Concluding remarks

The analysis of emissions from atomic and ionic species in laser induced plasma of metal copper shows dependence on the polarization state of the excitation laser light. Changes in emission line intensities of metallic Cu have been monitored by inducing changes in polarization direction of linearly polarized laser beam using a half-wave plate, as well as by conversion of polarization state from linearly polarized to circularly polarized using a quarter-wave plate. The emission characteristics observed are different for atomic and ionic emissions when the excitation beam is linearly and circularly polarized. Interesting results are seen in the case of ionic emission lines when different polarization directions of the laser beam were used for excitation and the enhanced emission is obtained when the polarization angle is around $45^0$ and $135^0$. The changes in emission characteristics are tentatively evaluated in terms of surface plasmon polariton formation and changes in radial potential functions of atoms with different polarization states of excitation laser beam. Our experimental observations are of immediate utility in LIBS in making use of the polarization properties of the excitation laser in order to optimize emission lines from both atomic and ionic transitions. Our findings also highlight the urgent need for appropriate theoretical to be undertaken so that proper insights into the physics that drive the observations that we report here.

## 5. Acknowledgements

We gratefully acknowledge financial support from the Device Development Program, Department of Science & Technology (DST), Government of India (DST/TDT/DDP-26/2018) and Department of Atomic Energy (DAE), Board of Research in Nuclear Sciences (BRNS), Government of India (34/14/04/2014-BRNS). Also, acknowledge support through the DST-Fund for Improvement of16

S&T Infrastructure (FIST) program (SR/FST/PSI-174/2012). U. K. Adarsh is thankful to Manipal Academy of Higher Education (MAHE) for the research fellowship provided## 6. References

1. Darwiche. S, Benmansour. M, Eliezer. N, and Morvan. D., *Investigation of optimized experimental parameters including laser wavelength for boron measurement in photovoltaic grade silicon using laser-induced breakdown spectroscopy.* Spectrochimica Acta Part B: Atomic Spectroscopy, 2010. **65**(8): p. 738-743.
2. Lasheras. R., Bello-Gálvez. C, and Anzano J., *Identification of polymers by libs using methods of correlation and normalized coordinates.* Polymer testing, 2010. **29**(8): p. 1057-1064.
3. Unnikrishnan. V.K, Rajesh. N, Aithal K., Kartha. VB, Santhosh. C, Gupta. GP, Suri. BM., *Analysis of trace elements in complex matrices (soil) by Laser Induced Breakdown Spectroscopy (LIBS).* Analytical Methods, 2013. **5**(5): p. 1294-1300.
4. Afgan M.S, Hou.Z, and Wang Z., *Quantitative analysis of common elements in steel using a handheld μ-LIBS instrument.* Journal of Analytical Atomic Spectrometry, 2017. **32**(10): p. 1905-1915.
5. Cremers D.A., Multari R.A, and Knight A.K., *Laser-induced breakdown spectroscopy.* Encyclopedia of Analytical Chemistry: Applications, Theory and Instrumentation, 2006: p. 1-28.
6. Miziolek A.W, Palleschi V, and Schechter I., *Laser induced breakdown spectroscopy.* 2006: Cambridge university press.
7. Senesi G.S, Manzari P, Consiglio A, and De Pascale O., *Identification and classification of meteorites using a handheld LIBS instrument coupled with a fuzzy logic-based method.* Journal of Analytical Atomic Spectrometry, 2018. **33**(10): p. 1664-1675.
8. Unnikrishnan V.K, Rajesh N, Praveen D, Tamboli M.M, Santhosh C, Kumar G.A, and Sardar D.K., *Calibration based laser-induced breakdown spectroscopy (LIBS) for quantitative analysis of doped rare earth elements in phosphors.* Materials Letters, 2013. **107**: p. 322-324.
9. Shameem K.M.M, Choudhary K.S, Aseefhali B, Kulkarni S.D, Unnikrishnan V.K, Sajan D.G, Kartha V.B, and Santhosh C., *A hybrid LIBS–Raman system combined with chemometrics: an efficient tool for plastic identification and sorting.* Analytical and bioanalytical chemistry, 2017. **409**(13): p. 3299-3308.
10. Liu K, Tian D, Wang H, and Yang G, *Rapid classification of plastics by laser-induced breakdown spectroscopy (LIBS) coupled with partial least squares discrimination analysis based on variable importance (VI-PLS-DA).* Analytical methods, 2019. **11**(9): p. 1174-1179.
11. Stratis D.N., Eland K.L, and Angel S.M., *Dual-pulse LIBS using a pre-ablation spark for enhanced ablation and emission.* Applied Spectroscopy, 2000. **54**(9): p. 1270-1274.
12. Popov A.M., Colao F, and Fantoni R., *Enhancement of LIBS signal by spatially confining the laser-induced plasma.* Journal of Analytical Atomic Spectrometry, 2009. **24**(5): p. 602-604.
17